%% file: main.tex
\documentclass{stsci_report}

\usepackage[backend=biber]{biblatex}
\usepackage{graphicx}
\usepackage{siunitx}
\usepackage{todonotes}
\usepackage{pdfpages}
\usepackage{subcaption}
\usepackage{scalerel}
\usepackage{tikz}
\usetikzlibrary{svg.path}
\usepackage{amsmath,amssymb}
\usepackage{wrapfig}
\usepackage{multicol}
\usepackage{mdframed}
\usepackage[hyperfootnotes=false]{hyperref}
\usepackage{tabularx}
\usepackage{multirow}
\usepackage[indent]{parskip}
\usepackage{indentfirst}
\usepackage{multirow}
\usepackage{longtable}
\usepackage{caption}
\usepackage[bottom]{footmisc}
\usepackage{wrapfig}
\usepackage{amsmath}
\usepackage{colortbl}
\usepackage{tabularray}
\usepackage{setspace}
\usepackage{lscape}
\usepackage{comment}

\copyrighttext{Copyright\copyright\ \the\year\ The Association of Universities for Research in Astronomy, Inc. All Rights Reserved.}

\presubtitle{Instrument Science Report WFC3 2025-03} 

\title{WFC3/IR Starter Guide}
\author{P. R. McCullough, Joel D. Green (corresponding author)}
\date{July 15, 2025}

\begin{document}

\include{abbreviations}

\maketitle

\abstract{In this starter guide, we provide a high-level overview of analysis of WFC3/IR data available from the Mikulski Archive for Space Telescopes (MAST). We intend this guide as a starting point for users examining WFC3/IR data for the first time, or for those refreshing their memory on WFC3/IR data analysis. Therefore, we focus on the analysis of archival data, not preparing new observations. Three appendices include A) a summary of the instrument and an optical schematic, B) examples from the Exposure Time Calculator, and C) a glossary of uncommon acronyms. This report addresses only data from WFC3's IR channel; \underbar{not} the UVIS channel.}

\newpage
\tableofcontents
\newpage

\section{Introduction} \label{sec:intro}

\begin{figure}[h]
\centering
\includegraphics[width=0.96\textwidth]{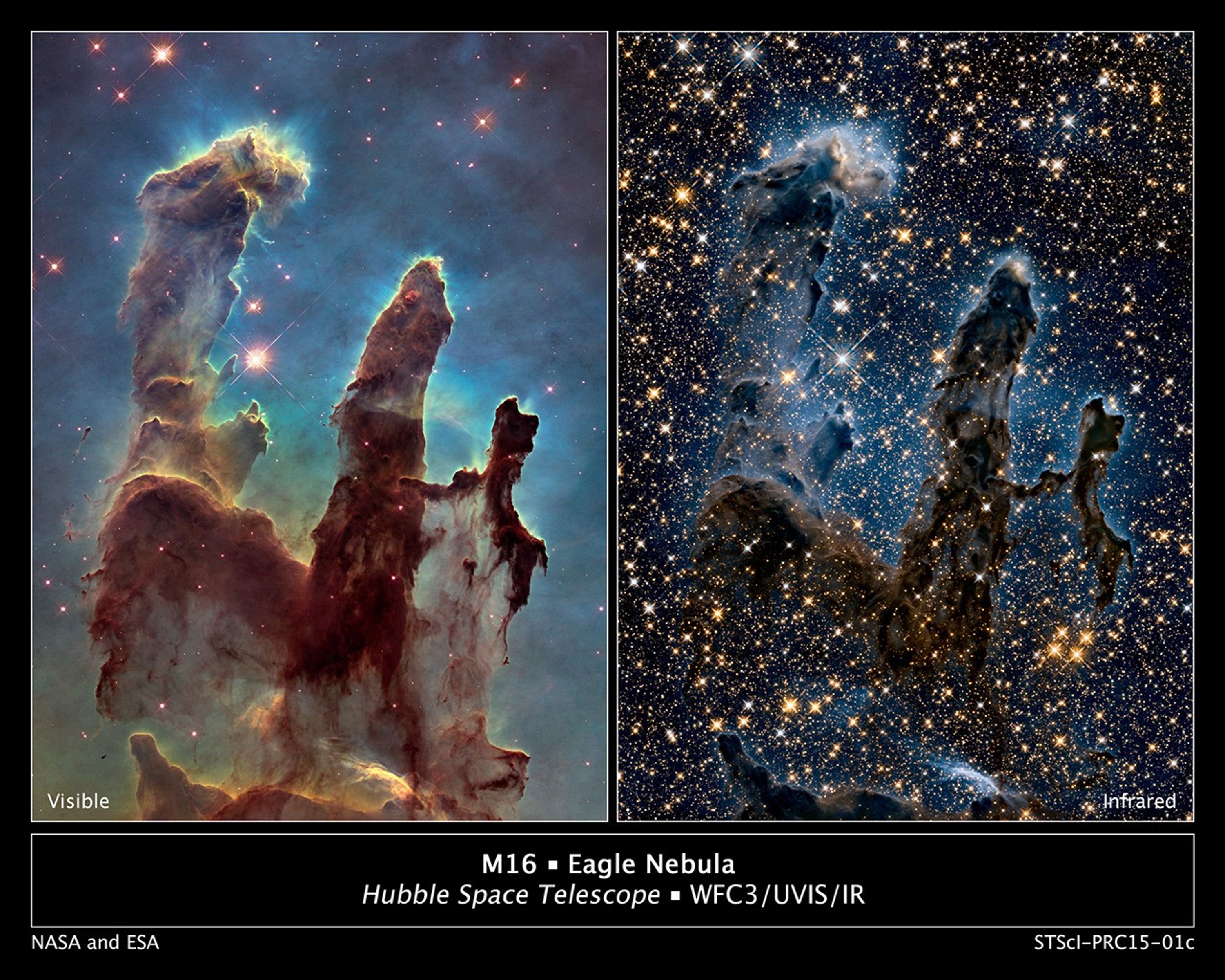}
\caption{{\bf Left:} WFC3/UVIS mosaic of the ``Pillars of Creation,'' part of the Eagle Nebula. {\bf Right:} WFC3/IR mosaic of the same field of view. Credit: NASA/ESA/STScI.}
\label{fig:pillars}
\end{figure}

Installed on the Hubble Space Telescope (HST) in 2009, the Wide Field Camera 3 (WFC3) included two channels: UVIS (UV and visible) and IR (infrared). WFC3/IR improved HST's performance with 20$\times$ greater sensitivity over the previous IR instrument, NICMOS (Near Infrared Camera and Multi-Object Spectrometer). Combined with HST’s spatial resolution and sensitivity, WFC3/IR created entirely new opportunities for astrophysical research. 

WFC3/IR also provided some of the best support for the development of future IR missions, in combination with the Spitzer Space Telescope. Unique visuals such as the Pillars of Creation (Figure \ref{fig:pillars}), with their starkly different appearance in IR, motivated high-resolution and mid-IR instruments on the James Webb Space Telescope (JWST). The detectors on JWST's Near Infrared Camera (NIRCam), for example, are direct descendants of those flown on WFC3/IR. 

As of this publication in 2025, WFC3/IR continues to play a vital role in NASA's suite of space missions. Although some of its capabilities overlap with JWST's NIRCam, WFC3/IR remains a scientifically powerful instrument on HST. However, WFC3/IR support may decrease over time, with a corresponding reduction in available expertise for users. With that in mind, this document aims to help preserve the legacy and usability of the WFC3/IR archive, which contains a wealth of imagery and spectra of solar system planets, exoplanets, stars, molecular clouds, supernovae, the local group, nearby galaxies, large-scale structure, and the distant universe, observed in numerous modes from hundreds of innovative scientific programs selected by a highly competitive peer review process. The purpose of this document is to assist users to make discoveries long into the future. 

This guide is intended as a ``first stop'' for new users of WFC3/IR, or for experienced users in search of a refresher. {\bf It is not an exhaustive manual}, but rather an abbreviated guide summarizing tips, tricks, references, and links for further study. In this guide, we provide a series of steps to acquire, view, and analyze data. In addition, we highlight each observing mode and list a number of characteristics and properties associated with WFC3/IR data.

The key documents on the
WFC3 website\footnote{\url{https://www.stsci.edu/hst/instrumentation/wfc3}} are as follows:
\begin{itemize}
  \item The Instrument Handbook, {\bf IHB}, \textcite{ihb}, 
  a 700 page document describing the instrument properties, status, and capabilities, intended primarily for observers.
  \item The Data Handbook, {\bf DHB}, \textcite{dhb}, 
  a 300 page document describing WFC3 data structure, integrity, and analysis, intended for primarily for analysis of archival data.
  \item Instrument Science Reports, {\bf WFC3-YYYY-nn}, generally 10-20 page in-depth studies of a single calibration characteristic of the instrument.
\end{itemize}
In addition, the following websites are especially useful: 
\begin{itemize}
  \item The MAST search engine or portal (\S\ref{sec:quick})
  \item A library of Python Jupyter Notebooks for data analysis (\S\ref{sec:software})
  \item The DrizzlePac notebooks for combining images\footnote{\url{https://www.stsci.edu/scientific-community/software/drizzlepac}} 
\end{itemize}

\textsf{Note to the reader: any paragraph or more of prose in this report that has been copied verbatim, or nearly so, from another STScI document appears in sans-serif font (as does this paragraph) and is accompanied by a trailing citation to the original source. We have sought and secured permission from the corresponding authors. No such reproduction occurs from any source outside STScI. For the sake of brevity, references that appear within such text generally do not appear in this document; instead, check the original source. The abbreviations \fnihb, \fndhb, or WFC3-YYYY-nn are used for those specific citations. 
}

\section{The Main Steps, from Download to Science} \label{sec:quick}

\noindent {\bf Download}

Users may download the data from MAST using a) \verb|astroquery| package within \verb|astropy|\footnote{\url{https://astroquery.readthedocs.io/en/latest/mast/mast.html}}, b) the HST MAST Search Engine\footnote{\url{https://mast.stsci.edu/search/ui/\#/hst}}, or c) the more general MAST Portal\footnote{\url{https://mast.stsci.edu/portal/Mashup/Clients/Mast/Portal.html}}. The HST search engine is best for individual queries and finding specific datasets. The Portal is a GUI that enables advanced searches of HST-specific combinations of instrument, proposal number, filter names, and other parameters. It too is searchable by many different parameters (e.g., RA/Dec, release date, filter, program ID, etc.), and contains data from many missions including HST, JWST, Kepler, GALEX, TESS, FUSE, and others.  

The most critical files for analysis are the Intermediate MultiAccum (IMA) files (\texttt{\_ima.fits}), which contain the full 3D cube of calibrated readouts. We highly recommend all IMA reads be examined, especially for data taken with the F105W or F110W filters (explained in more detail below). The FLT (\texttt{\_flt.fits}) image is a single 2D image derived by the calwf3 pipeline's fitting up-the-ramp of the accumulated charge in each pixel stored in the corresponding IMA file. The IMA files can be used to verify any features suspected to be instrumental or spurious, which can be obscured by the additional processing steps that create the FLT files.  If available for a given observation, one can also download the HAP (Hubble Advanced Products), which are single- or multi-visit mosaics of images co-aligned and combined by drizzling (cf. below). HAP files are convenient, but they have known caveats for photometry and astrometry that can be corrected with expert scientific analysis (cf. WFC3-2022-06).

\noindent {\bf Display}

Many astronomers use the data visualizer \texttt{ds9} developed by the Smithsonian Astrophysical Observatory (SAO) to visually inspect 2D images such as FLT files or 3D image cubes such as IMA files. In addition, for display and manipulation of FITS files, the user may have their own favorite general-purpose software, e.g. the proprietary Interactive Data Language (IDL) or the public-domain Python libraries such as \texttt{astropy} and its \texttt{display\_image} which will display the science (SCI), error (ERR), and data quality (DQ) array for an image (\S \ref{sec:dataproducts}). For more information regarding file structure, consult Chapter 2 of the DHB.

Drizzle (DRZ) products (\texttt{\_drz.fits}) are images generated by MAST that are corrected for alignment, distortion, cosmic rays, and combined together onto a common World Coordinate System (WCS) updated by aligning sources in the field of view to an external reference catalog (e.g., Gaia). These products can serve as an excellent starting point for many analyses. For example, DRZ products are excellent for making quality extended mosaics and overlays with other images. However, the MAST automated products may not optimally align the constituent images. Thus, users interested in precision astrometry and/or fine-tuning their mosaics should consider drizzling their images themselves (WFC3-2022-06; DrizzlePac python notebooks).

\noindent {\bf Inspect}

Inspect the data, investigating the data quality and checking for idiosyncrasies. The most common image artifacts of the WFC3/IR data are cosmic rays, persistence, dead pixels, blobs, and occasionally a time-variable background. These are discussed in \S \ref{sec:artifacts}. After accounting for these effects, the data should be ready for science!

\section{Data Visualization and Analysis}
\label{sec:datavis}

Data obtained from MAST, calibrated by the nominal CALWF3 pipeline, are summarized in \S \ref{sec:dataproducts}. This section summarizes how to visually inspect the data files obtained from MAST, and also notes some additional data reduction that may be necessary prior to scientific analysis (cf. \S \ref{sec:software}).

\subsection{Data Artifacts}
\label{sec:artifacts}

In this section, we show how to quickly identify commonly encountered anomalies in IR data and how to remedy each. Many artifacts are more visible in the IMAs than in the FLTs, because the up-the-ramp fitting tends to hide artifacts. New users are encouraged to consult \S \ref{sec:dataproducts} for nomenclature and \S \ref{sec:glossary} for acronyms. Error sources for WFC3/IR are detailed in Chapter 7 of the DHB.

The sensitivity of WFC3/IR has been declining at ${\sim}0.1$\% per year, although it depends on wavelength, with slower declines in the redder filters (WFC3-2024-06). The FITS keyword \texttt{PHOTFLAM} used for conversion to absolute units (e.g. erg s$^{-1}$cm$^{-2}$ per \AA) includes the appropriate time dependence (DHB Ch. 9). Note that the time-dependent correction is encoded in the keyword only; the conversion to absolute units is NOT automatically applied to the science data arrays. Thus, users who wish to compare results from different epochs and/or stack data from different epochs will need to apply this conversion to the data arrays first.

\begin{figure}[h]
\centering
\includegraphics[width=0.37\textwidth]{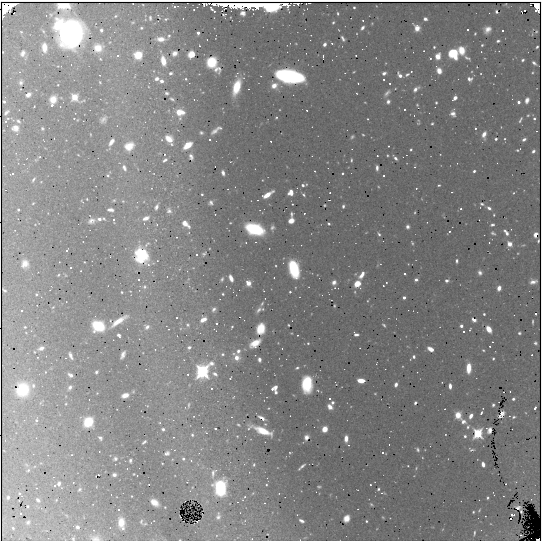}
\includegraphics[width=0.37\textwidth]{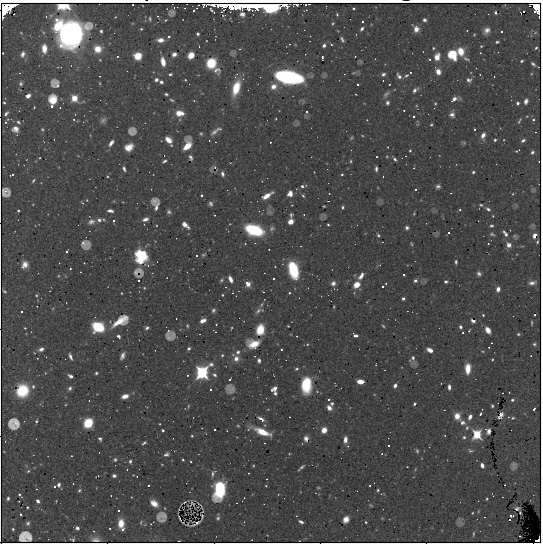}
\\
\includegraphics[width=0.37\textwidth]{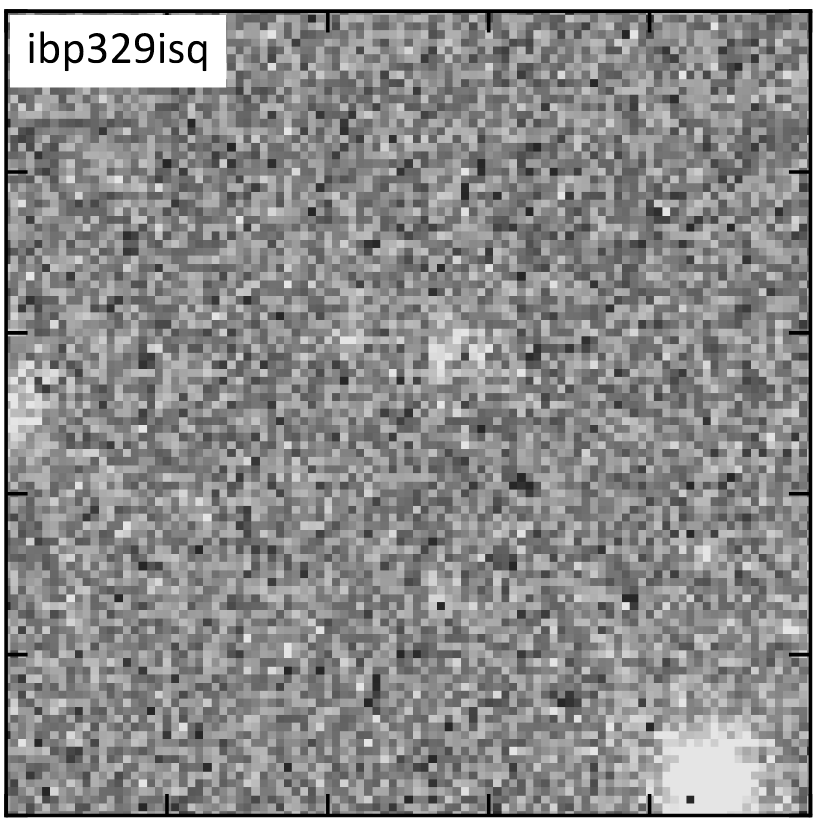}
\includegraphics[width=0.37\textwidth]{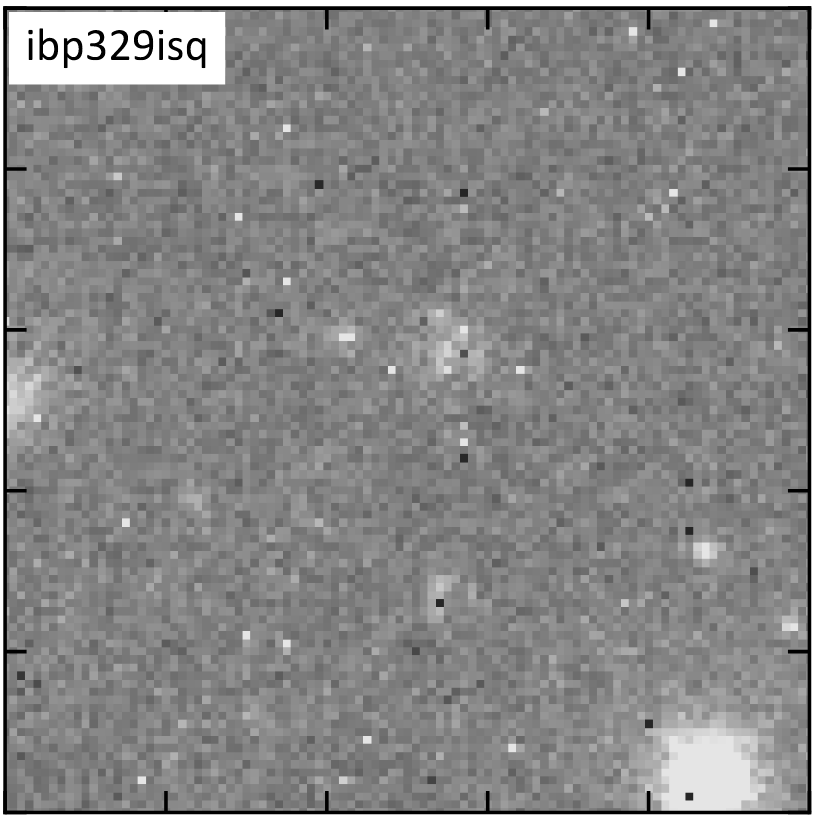}
\caption{One example of Time-Variable Background (TVB) contamination before (left) and after manual correction (right). The top row illustrates the full field of view; scattered Earthshine is characteristically brighter (whiter) in the left ${\sim}200$ columns of WFC3/IR images than the rest of the image. The bottom row zooms in on a 200×200 pixel region, providing detail and illustrating the improvement achieved by digital removal of TVB (cf. \S \ref{sec:software}) \fndhb.}
\label{fig:tvb}
\end{figure}

\textsf{{\bf Time-variable background (TVB)} is an artifact not corrected by the CALWF3 pipeline in the default calibrated images from MAST (Figure \ref{fig:tvb}). The IR background for HST is a combination of zodiacal light, scattered light from the bright Earth limb, and line emission at 1083 nm from helium atoms in the exosphere, through which HST is orbiting. Zodiacal light is essentially constant spatially and temporally within an exposure and an HST orbit. However, the scattered light and line emission components can vary temporally within an orbit and even within a single exposure. The scattered light varies in a characteristic pattern across the field of view, whereas the line emission does not (Figure \ref{fig:tvb}). The scattered light has a solar spectrum, so it can affect both grisms and all the filters. The helium line is transmitted by both grisms, especially the G102 grism (see \S \ref{sec:spectroscopy}), but only by specific filters (F105W, F110W, and F098M). \fndhb}

\textsf{Strong time variation in the background (\S \ref{sec:artifacts}) during a MULTIACCUM ramp can confuse the pipeline's cosmic ray identification algorithm CRCORR, which assumes that each pixel experiences its own particular count rate and that any large deviations from that rate are due cosmic-ray hits (not TVB). Furthermore, the variationin the background caused by the helium line is frequently not obvious in FLT files, requiring close evaluation of IMA files.
The confusion can cause most or nearly all pixels in an image to be misidentified as cosmic ray hits in many of the samples up the ramp, \underbar{markedly} increasing the noise in the image (Figure \ref{fig:tvb}). As a consequence, MAST-generated \texttt{\_flt.fits} images that have TVB are not recommended for scientific analysis and instead should be manually reprocessed (\S \ref{sec:software}). \fndhb}

{\bf Cosmic rays} appear as long and narrow streaks of hot or saturated pixels, or sometimes as more rounded blob patterns (referred to as ``snowballs'').  They are typically removed by the CALWF3 pipeline in the IMA-to-FLT processing, 
by comparing multiple samples in the ramp. However, cosmic ray hits can be missed if there are too few samples up the ramp of accumulating charge, e.g. NSAMP$\lesssim 7$. Python notebooks are available for correcting the IMA files for TVB (\S\ref{sec:software}).

{\bf Persistence} is an electronic afterglow of earlier images, particularly from sources that fill more than half of the full-well of a pixel. This filling could have occurred up to a few days prior, or from data acquired within the same HST orbit. Pixels near popular locations (e.g. near the center of an array) are more likely to be affected, but any part of the detector can be affected. 
If persistence is suspected, check for very bright sources in previous observations, either by a MAST search, or by using the WFC3/IR persistence websites\footnote{\url{https://archive.stsci.edu/prepds/persist/search.php} and \url{https://www.stsci.edu/hst/instrumentation/wfc3/data-analysis/ir-persistence}}.

\begin{figure}[h]
\centering
\includegraphics[width=0.48\textwidth]{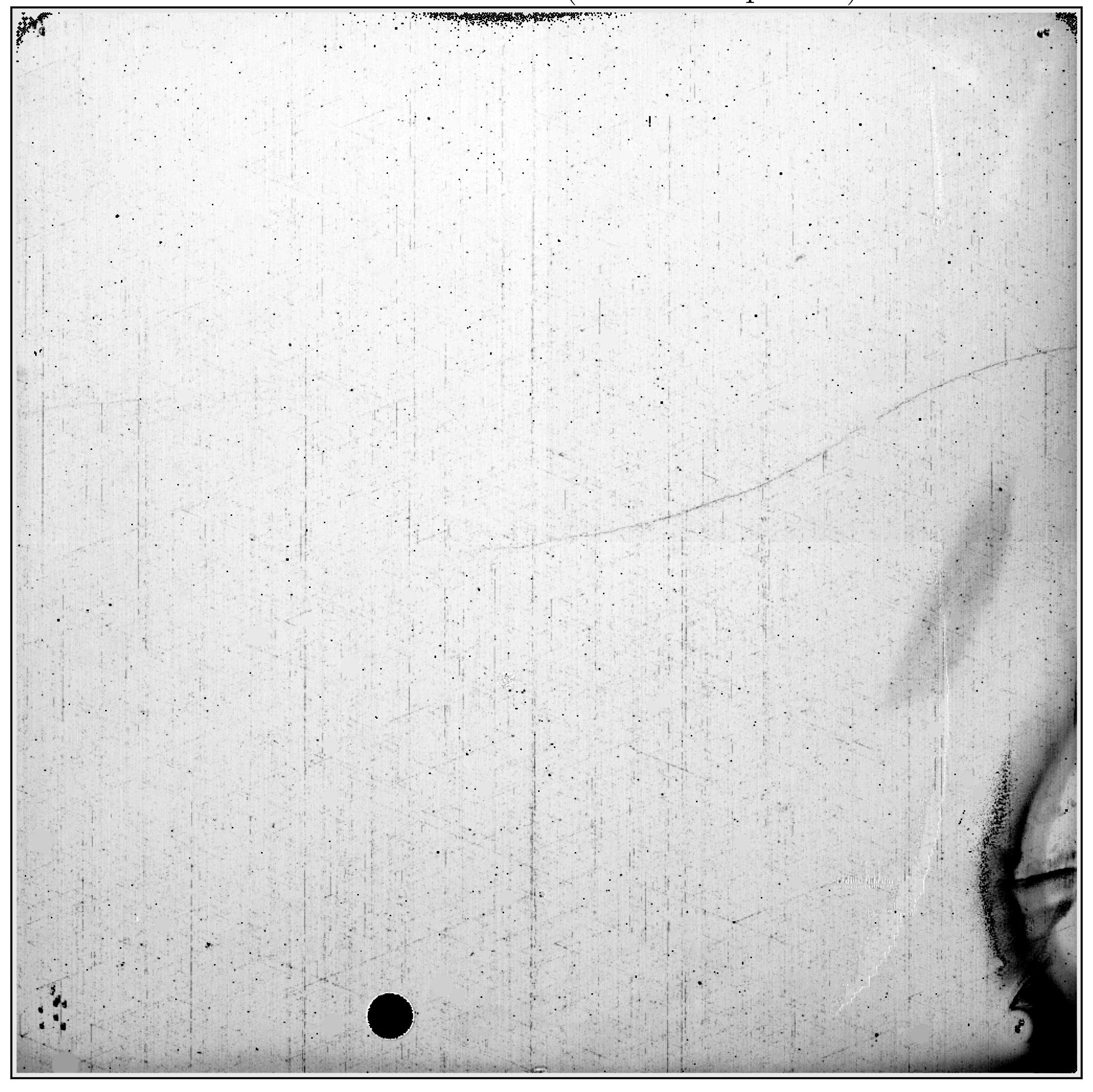}
\includegraphics[width=0.48\textwidth]{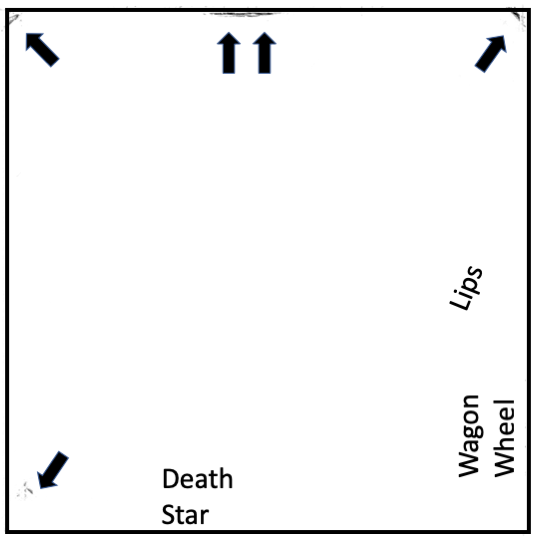}
\caption{{\bf Left:} F110W internal flatfield image (``p-flat'') 4ac1921ri\_pfl.fits (from IHB). {\bf Right:} clusters of dead or anomalous pixels (arrowed) and three named features.}
\label{fig:deadpixels}
\end{figure}

{\bf Dead pixels} and other features (with imaginative nicknames) are best seen in the flat field (Figure \ref{fig:deadpixels}, also see DHB Ch. 8). Prominent clusters of dead pixels occur near the edges and corners (cf. arrows in Figure \ref{fig:deadpixels}). The ``Death Star'' is a unique circular patch of dead pixels, which is helpful in establishing the detector's orientation in a manipulated image (e.g. rotated or flipped). Similarly, dead pixels can be useful as reseaux (fiducial) marks to establish precise registration in user-created software. The ``wagon wheel'' is a region with anomalous quantum efficiency (larger variation and color-dependent sensitivity) that may produce less reliable data. The ``lips'' is another named region, but it produces reliable data.

{\bf IR blobs} are circular areas a few pixels in diameter with a few percent lower sensitivity, caused by bits of debris stuck to the channel select mechanism's mirror (WFC3-2010-06). Their number has increased monotonically over the lifetime of the mission. As of mid-2025, there were about 50 blobs affecting 12891 pixels, or about 1.2\% of the array, that are flagged in the bad pixel table and propagated into the science data quality file. (Another $\sim$ 100 blobs were not included because they were small and/or faint enough to drop out with stacking of dithered exposures.) Blobs are in sharp focus (${\sim}1$ pixel across) in the upper right quadrant, but are out of focus (a few pixels across) in the lower left quadrant (the one with the Death Star). In direct images through F098M or any of the five wide filters, blobs that appeared through 2018 are are corrected by an appropriate delta-flat field (WFC3-2021-10); blobs after 2018 may require manual correction by the user.  Although the blob regions are identified by bit number 9 in the DQ arrays, MAST ignores that specific bit ($2^9$ = 512) to create its nominal drizzled data products. However, users may wish to apply it to images dithered with the WFC3-IR-DITHER-BLOB pattern.

\subsection{Initial Analysis for Specific Science Goals}

Once the data files are downloaded and inspected, preparing them for scientific analysis depends on the specific goals, and will require some careful consideration. 

Stellar photometry can be obtained directly from drizzled DRZ files in which the pixel areas on the sky are invariant across the field of view. Alternatively, to perform aperture photometry on FLT files, one must correctly apply a pixel area map (PAM) to compensate for the variation in pixel areas across the field of view. 

Photometric FITS keywords (e.g. PHOTFLAM) enable conversion to absolute flux units (erg s$^{-1}$cm$^{-2}$ per \AA\ or Jy str$^{-1}$, or magnitude units in the STMAG, ABMAG, or VEGAMAG systems (cf. the DHB \S 9.1 and the Flux Conversion notebook in \S 6). 
For precise differential photometry over a large dynamic range, additionally apply the countrate non-linearity correction of 0.75\% per factor of ten in count rate to mitigate reciprocity failure (WFC3-2019-01). 

For photometry, it is important to note that the IR PSF is undersampled. PSF analyses should make use of the model PSFs available on the WFC3 pages or custom-built PSFs (see Notebook Repository for cookbook). The PSF library described in WFC3-2021-12 generally will produce superior results to \texttt{TinyTim}\footnote{TinyTim is not supported and should not be used for WFC3. While it employs some preliminary optical models, these have not been updated to reflect in-flight performance.} models.  A comprehensive PSF modeling notebook is available on the Notebook Repository\footnote{\url{https://spacetelescope.github.io/hst_notebooks/notebooks/WFC3/point_spread_function/hst_point_spread_function.html}}. The \texttt{hst1pass} routine can perform PSF-fitting or aperture photometry on WFC3/IR images (WFC3-2022-05).

\subsection{Error Budget} \label{sec:error}

\begin{figure}[h]
\centering
\includegraphics[width=0.8\textwidth]{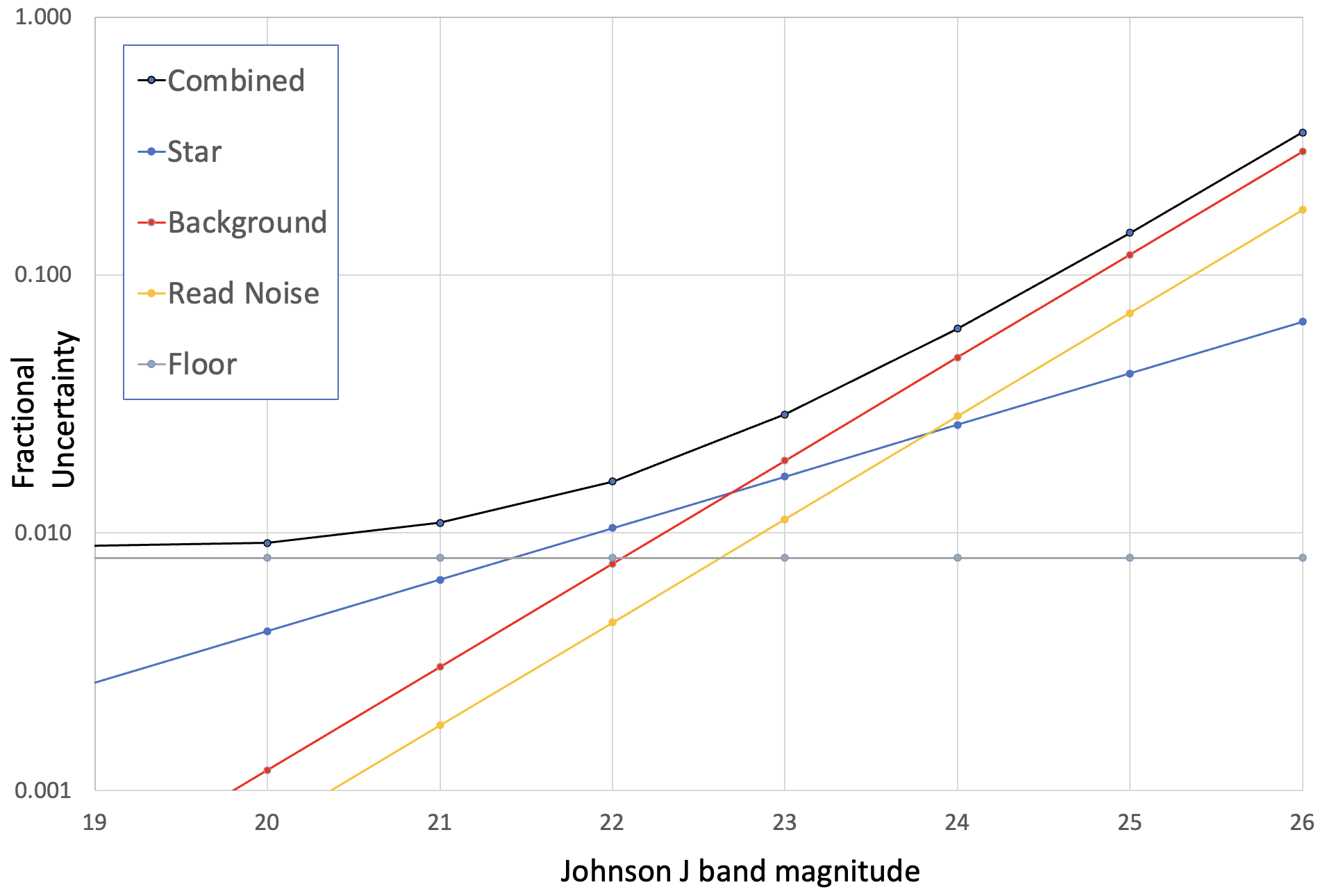}
\caption{The predicted fractional photometric uncertainty, ${\rm \sigma = 1/SNR}$, in dimensionless units, of a WFC3/IR F110W 353s observation of a solar type star (G2 V) plotted with respect to the star's J-band magnitude in the Vega photometric system. The combined value (black) is the quadrature sum of four components: A) Poisson noise from the star itself (blue); B) background noise (red); C) read noise (yellow), and D) a noise floor (grey). Results (points) are from the Exposure Time Calculator; e.g. the fractional uncertainty is 10\%, i.e. ${\sigma}=0.1$ or ${\rm SNR}=10$, at J=24.5 mag (see text for details).} 
\label{fig:error_vs_mag}
\end{figure}

{\bf Photometric repeatability} has four main components: A) the Poisson noise from the star itself, B) background noise from various sources: sky emissions from zodiacal light and Earthshine, thermal emission from the instrument and optics, and dark current from the detector; C) the read noise of the detector, and D) a systematic noise floor. Each of those can be estimated with the Exposure Time Calculator (ETC); Appendix B provides examples for many configurations.\footnote{We recommend replicating an example or two with the online ETC in order to be confident in understanding the many input parameters.} Figure \ref{fig:error_vs_mag} illustrates ETC predictions for one configuration: the F110W filter, a 353 s exposure (SPARS25, NSAMP=15), a 0.2 arcsec radius photometric aperture (7.66 pixel area), average values of Zodiacal emission and Earthshine, and an equivalent read noise of 15 e$^-$ r.m.s. per pixel. The ETC predictions\footnote{The ETC results are archived at \url{https://etc.stsci.edu/etc/results/WFC3IR.im.1983178/}} at J = 24 mag are, respectively, integrated over the aperture: A) the star contributes 1457 \electron, B) the sky contributes 4567 \electron, thermal 146 \electron, dark current 141 \electron, C) read noise is 41.3 \electron\ r.m.s., and D) the noise floor is zero (an idealization of the ETC, but for analysis we recommend a non-zero value\footnote{Observations exhibit a photometric noise floor of $\sim$0.8\% in staring mode (WFC3-2019-07) and $\sim$0.5\% in scanning mode (WFC3-2021-05). The main contributors to the systematic floor are errors in the nonlinearity correction, the time dependent IR sensitivity, the flat fields,  and charge persistence. Estimates of the systematic noise floor vary considerably depending on many details of the observational technique, analysis, and time scale of variation, which are beyond the scope of this guide. An expert analyst might extract more precise values than those reported here, but not typically.}). 

As such, the four noise contributions plotted at J = 24 mag in Figure \ref{fig:error_vs_mag} are A) 0.026, B) 0.048, C) 0.028, and D) 0.008; the combination of which equals 0.062, corresponding to SNR=16.13, i.e. 1/16.13 = 0.062. Aside from neglecting the noise floor, the ETC predictions typically have matched observational results within $\pm$10\%. Note that the straight lines in Figure \ref{fig:error_vs_mag} indicate power laws with respect to the stellar flux, i.e. the noise $\sigma \propto flux^\alpha $; the indices $\alpha$ of the contributors equal to A) -0.5, B) -1.0, C) -1.0, and D) 0.0, respectively. 

\textsf{The read noise equals 12 \electron\ r.m.s. for a 16-read linear fit, or $\sim$20 \electron\ r.m.s. for correlated double sampling. (The ETC assumes 15 \electron\ r.m.s.) The full well depth is $\sim$77,900 \electron\ and the dark current is 0.05 \electron s$^{-1}$ per pixel. \fnihb
}

The {\bf astrometric precision} of well-exposed stars in a single (staring-mode) WFC3/IR image can be $\sim$0.01 pixel  in either axis, X or Y, equivalent to $\sigma_x = \sigma_y \sim$0.001 arcsec. To achieve that, given the under-sampled PSF, advanced analysis is required, especially accounting for the sinusoidal offset with pixel phase, which has a peak-to-peak amplitude of $\sim$0.2 pixel (WFC3-2016-12).

\section{Quick Reference Guide for Instrument Modes}
The following subsections provide curated information for each instrumental mode. For full details, please see the IHB.

\subsection{Imaging}
\label{sec:imaging}

\textsf{The IR field of view is 136 arcsec by 123 arcsec, with each pixel covering 0.135×0.121 arcsec. The IR field of view is $\sim$3/4 the width of the UVIS field of view, and each IR pixel is $\sim$3x wider than a UVIS pixel. Only the inner 1014 × 1014 pixels are light-sensitive; the five reference pixels along each edge are not light-sensitive. As alternatives to the full frame, there are four {\bf(centered)} subarrays: 512×512, 256×256, 128×128, or 64×64. Including reference pixels, the data files are 1024x1024, 522×522, 266×266, 138×138, or 74×74 pixels. \fnihb
}

Table \ref{tab:limiting_magnitudes} lists the detection limits for direct imaging through two wideband filters (cf. Figure \ref{fig:error_vs_mag} and Appendix B). 

\begin{table}[h]
  \begin{minipage}{0.48\textwidth}
    \begin{tabular}{|c|c|c|c|}
        \hline
        Band & Filter & Mag. limit & Mag. limit \\ 
             &        & in 1 h & in 10 h \\  \hline
        J & F110W & 27.3 & 28.6 \\ \hline
        H & F160W & 26.6 & 27.9 \\ \hline
    \end{tabular}
    \caption{Limiting ABMAG magnitudes for direct images of stars at 10-$\sigma$ in different bands. \fnihb}
    \label{tab:limiting_magnitudes}
  \end{minipage}
  \hfill
  \begin{minipage}{0.48\textwidth}
  \begin{tabular}{|c|c|c|c|c|}
    \hline
    Grism & \multicolumn{3}{c|}{Spectral Type} & Wavelength \\ \cline{2-4}
      & O3 & A0 & G2 & \\ \hline
    G102 & 21.3 & 22.0 & 22.9 & 1050 nm \\ \hline
    G141 & 20.5 & 21.3 & 22.8 & 1550 nm \\ \hline
  \end{tabular}
  \caption{Each grism's limiting V-band magnitude at 5-$\sigma$ in 1h for main-sequence stars of three spectral types. \fnihb}
  \label{tab:grism_sensitivity}
  \end{minipage}
\end{table}

\begin{figure}[h]
  \centering
  \begin{minipage}{0.5\textwidth}
    \centering
    \includegraphics[width=\linewidth]{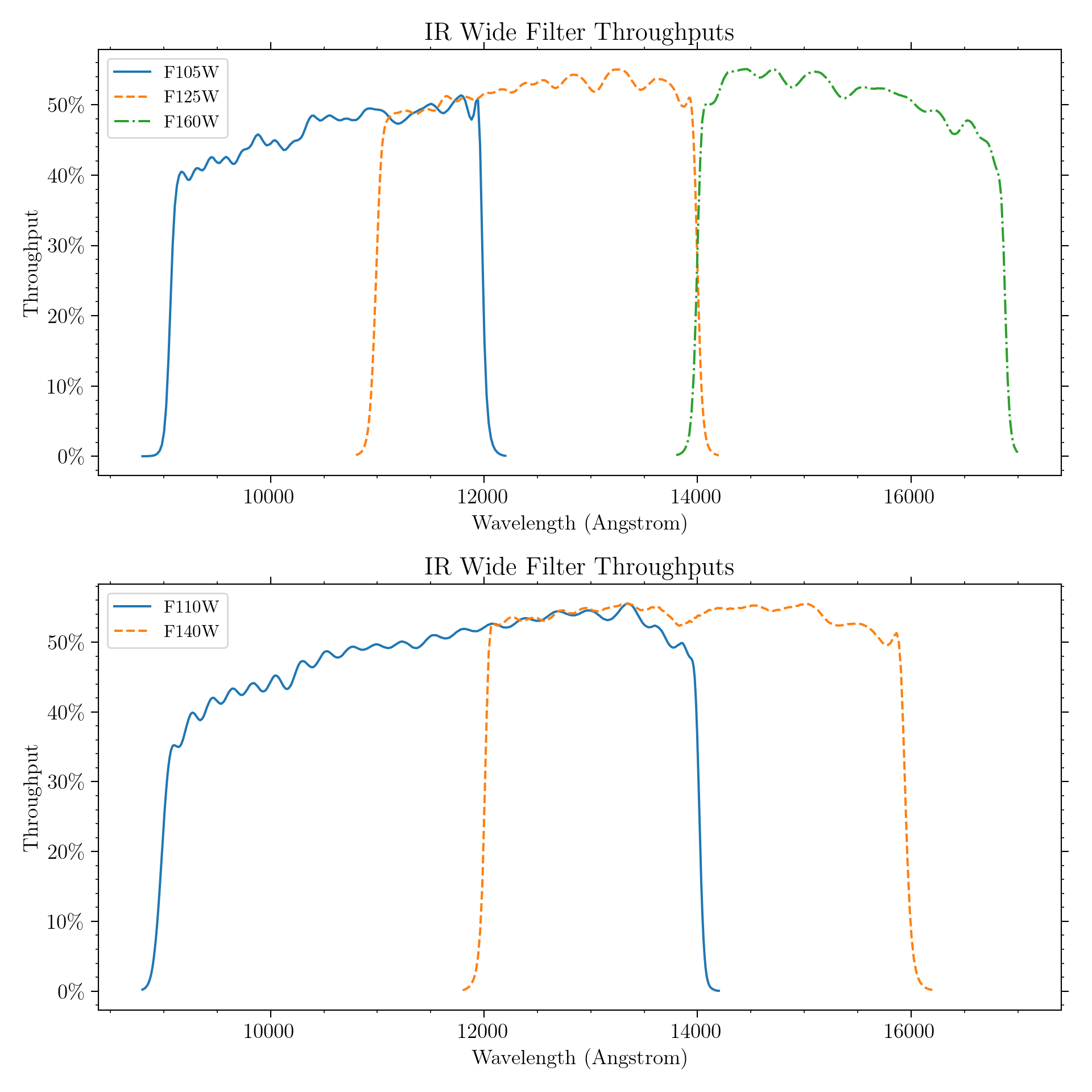}
  \end{minipage}\hfill
  \begin{minipage}{0.5\textwidth}
    \centering
    \includegraphics[width=\linewidth]{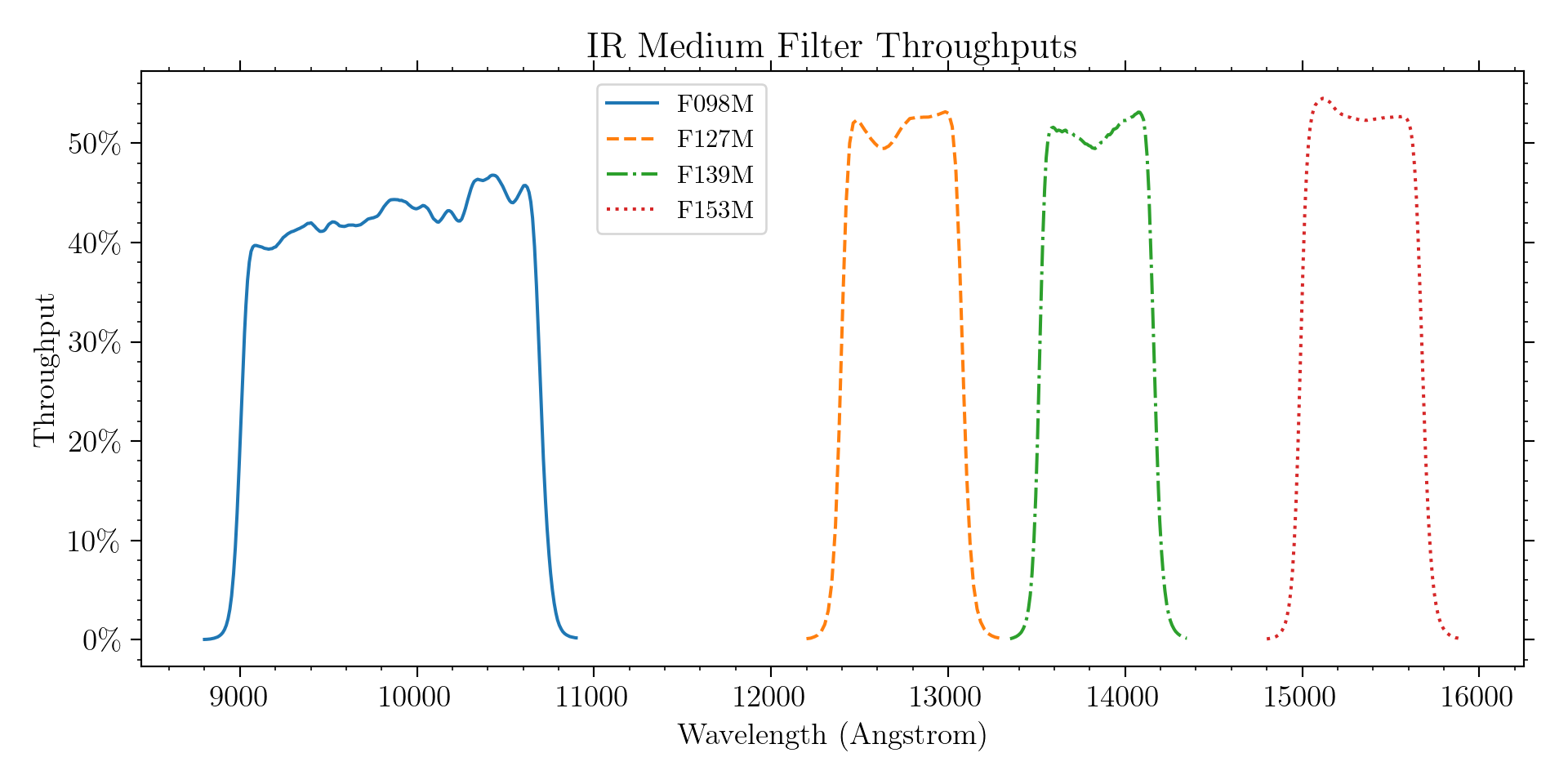}
    \includegraphics[width=\linewidth]{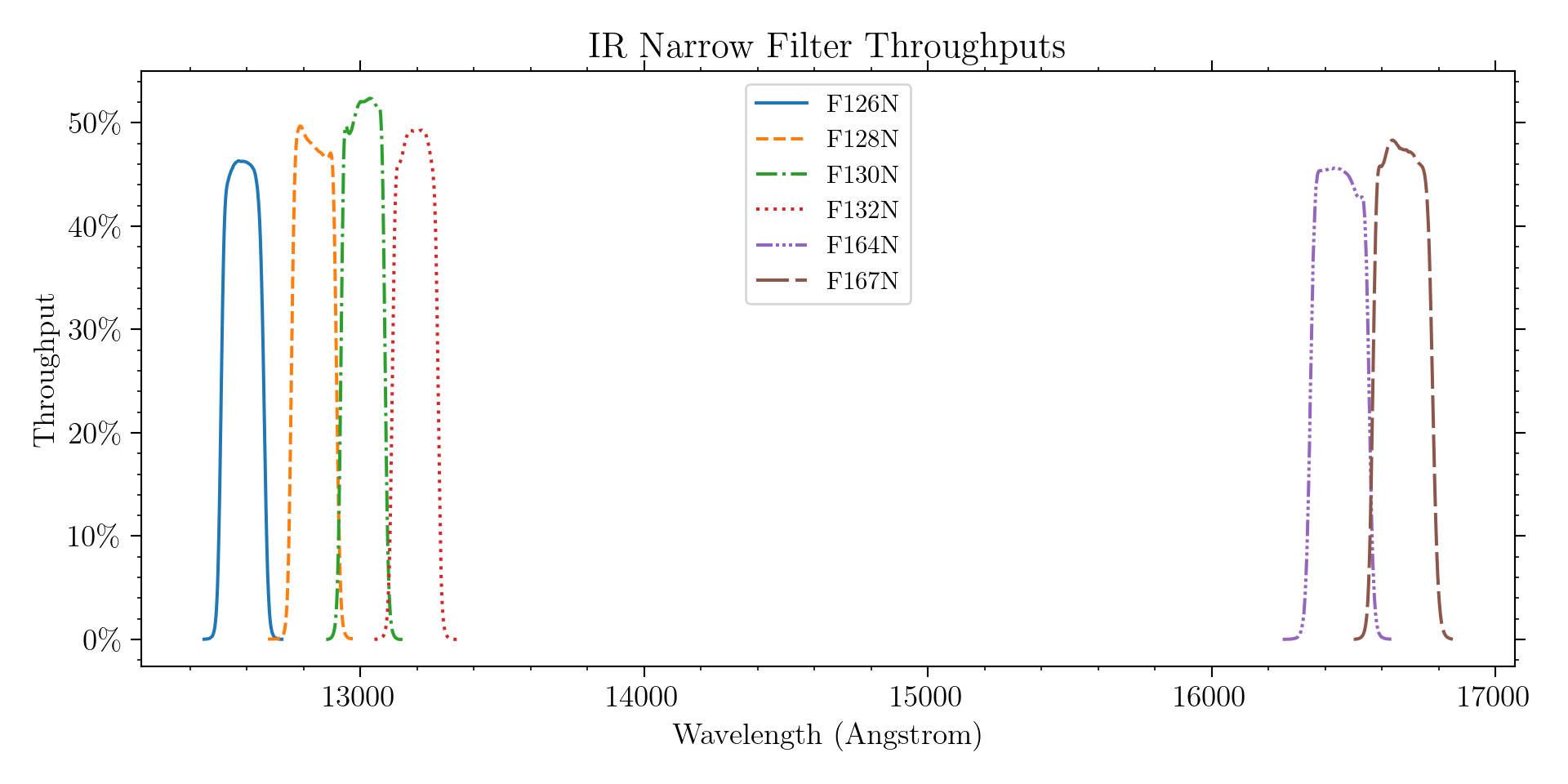}
  \end{minipage}
\caption{Filter throughputs (from the IHB).}
\label{fig:filters-wmn}
\end{figure}

Figure \ref{fig:filters-wmn} shows the throughputs of the filters: 5 wide, 4 medium, and 6 narrow. The values include the combined contributions of the telescope, the instrument, the filter, and the detector. The IHB contains additional
characteristics of the filters, including a table of pivot wavelengths, widths, and maximum throughput. Table 6.5 in the IHB quantifies out-of-band transmission, which is at most $\sim$1\% for only a few filters and less than 0.1\% for most of them, for objects with effective temperatures of 5000 K. (Stars hotter than 5000 K produce greater leaks.) 

\subsection{Spectroscopy}
\label{sec:spectroscopy}

\begin{figure}[h]
  \centering
  \begin{minipage}{0.5\textwidth}
    \centering
    \includegraphics[width=\linewidth]{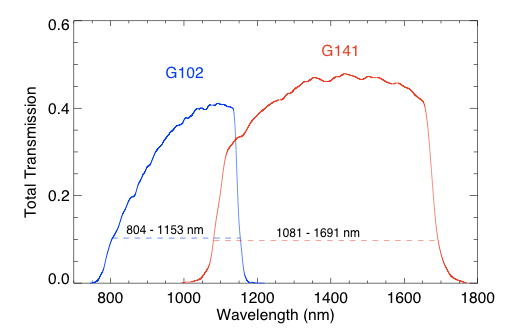}
  \end{minipage}\hfill
  \begin{minipage}{0.5\textwidth}
    \includegraphics[width=\textwidth]{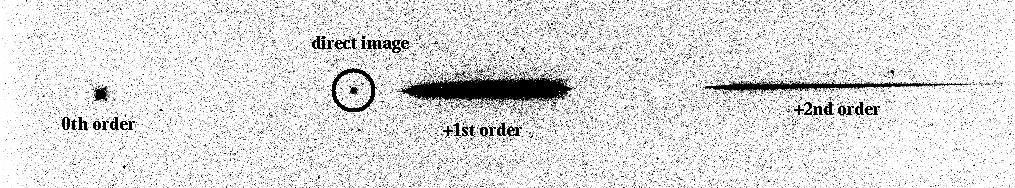}
    \\
    \\
    \includegraphics[width=\textwidth]{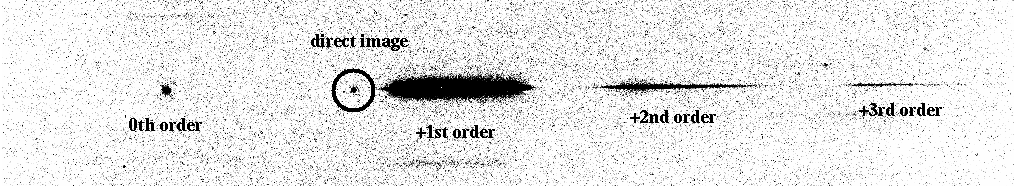}
    \\
  \end{minipage}
\caption{(Left) Grism throughputs in the +1 order. The spectral ranges with more than 10\% throughput are indicated. (Right) Grism images (G102 top, G141 bottom) of a star with a corresponding direct image (circled) superimposed at its nominal position relative to the grism image. Spectral orders are labeled. The image includes the full extent of the detector in the x-axis (1024 pixels) and about 200 pixels in the y-axis. \fnihb}
\label{fig:grisms}
\end{figure}

\textsf{There are two grisms for low-resolution slitless spectroscopy over the entire field of view of the detector (Figure \ref{fig:grisms}) They cover wavelength ranges of 800-1150 nm (G102) and 1075-1700 nm (G141) at first-order dispersions of 2.45 and 4.65 nm per pixel, and spectral resolution $\lambda/\delta\lambda$ of 210 and 130, respectively, given the PSF width of $\sim$2 pixels. \fnihb 
}

\textsf{Table \ref{tab:grism_sensitivity} lists the detection limits with the two grisms for three stellar spectral types. It tabulates the V magnitudes of stars for which a 1-hour exposure yields a spectrum with a S/N of 5 per resolution element, with an extraction box of 1 × 3 pixels. \fnihb
}

For most uses, grism data require special post-processing of the pipeline products (using e.g., HSTaxe or slitlessutils; \S\ref{sec:software}).

\subsection{Spatial Scanning} 
\label{sec:scan}

If desired during an exposure, HST can scan across the sky instead of the more common behavior of holding stars steady. 
\textsf{Spatially-scanned direct images can provide more precise photometry than staring mode can generally provide.}
Read \S 10.2 of the DHB before choosing to analyze either the \texttt{\_ima.fits} or the \texttt{\_flt.fits} files. 

\begin{figure}[h]
\centering
\includegraphics[width=9cm]{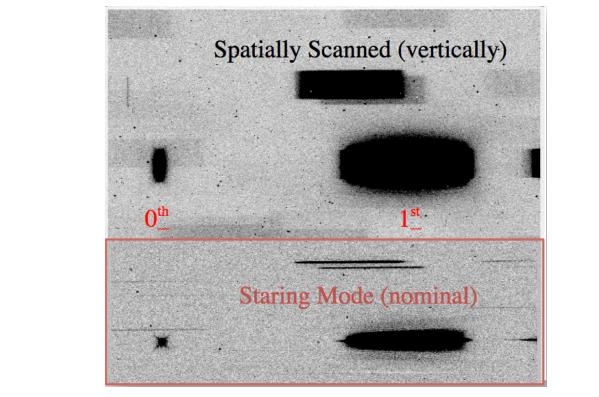}
\caption{A spatially-scanned G141 spectrum, labeled with its 0th and +1st order light, compared to a nominal staring-mode slitless spectrum of the same field (red outlined inset). The images are 512 columns wide, centered in the detector’s 1024 columns. The scan was 40 pixels high (4.8 arcsec). If a fainter star's spectrum overlaps (and contaminates) the target star's spectrum in the full scan, typically they can be separated by analyzing the individual reads of the sample sequence. \fnihb}
\label{fig:grisms_scanned}
\end{figure}

\textsf{Spatial scanning of stellar spectra creates the potential for spectrophotometry of higher precision than possible via staring mode. By spreading a stellar spectrum perpendicular to its dispersion (Figure \ref{fig:grisms_scanned}), more photons can be collected per exposure, and the exposure times can be longer without saturating the detector. This strategy results in higher signal-to-noise observations and an increased duty cycle. Be aware that if scans were performed in a forward-reverse ``round-trip'' manner, the flux may alternate up and down every other exposure, due to the electronic scanning of the detector. The most prevalent scientific application is transmission spectroscopy, in which a time series of stellar spectra are obtained before, during, and after an exoplanet transit or eclipse; observations of this type with a one-sigma precision of $\sim$20 ppm in $\sim$20 nm spectral bins from two transits have been reported (cf. citations in the IHB). \fnihb}

\subsection{Additional Modes} 
\label{sec:additionalmodes}

In addition to nominal modes of direct imaging and slitless spectroscopy, a few additional modes of operation have been used. 
\begin{enumerate}
  \item A Drift And SHift (DASH) observing technique enables IR mosaics without requiring guide stars (WFC3-2021-02).
  Although the telescope drifts under only gyroscopic control, special data analysis overcomes the blurring by extracting each very short exposure image from the IR sample sequence and then aligning and combining them. DASH observations are not noted by a specific keyword, but can be identified by the guidestar keyword T\_SGSTAR=N.A. Otherwise, drifts are unintentional and are not DASH mode.
  \item Moving targets (keyword MTFLAG=T) can be tracked in real time by HST, resulting in intentionally trailed background stars but a sharp target, e.g. a comet, asteroid, planet, natural satellite (a.k.a. moon), or even a specific feature on a rotating planet. 
\end{enumerate}

\section{Data Products and Calibration Steps}
\label{sec:dataproducts}

\textsf{All WFC3 science data products are two-dimensional images stored in Multi-Extension FITS format files. For each exposure taken with WFC3, there is one FITS file with a unique 9-character rootname followed by a 3-character suffix: rootname\_xxx.fits. The rootname uniquely identifies the observation, although superficially it looks like a pseudo-random string of alphanumerics. The suffix denotes the file type, e.g. RAW, IMA, or FLT. The IR channel operates only in MULTIACCUM mode, which starts an exposure by resetting all detector pixels to their bias levels and recording those levels in an initial ``zeroth'' readout. This is then followed by N non-destructive readouts (N can be up to 15 and is set by the observer as parameter NSAMP); the data associated with each readout are stored in a separate {\it imset} in the FITS file. Thus, the FITS file has N+1 imsets in total. \fndhb}

\textsf{For IR data, each imset consists of five data arrays:
\begin{enumerate}
  \item The science image (SCI),
  \item The error array (ERR),
  \item The data quality array (DQ),
  \item The number of samples array (SAMP), and
  \item The integration time array (TIME).
\end{enumerate}
The IMA (\texttt{\_ima.fits}) FITS file contains the primary header unit and N imsets, which together form a single IR exposure, and which results from the pipeline calibration of the RAW (\texttt{\_raw.fits}) FITS file. The FLT (\texttt{\_flt.fits}) FITS file contains only a single imset after the CRCORR ramp fitting of the data in the IMA file.
 \fndhb}

\textsf{The calibration steps are briefly summarized below, in the order they are executed, with the corresponding calibration switch keyword in parenthesis:
\begin{enumerate}
  \item Initialize data quality, DQ, array (DQICORR)
  \item Estimate amount of signal in zeroth-read (ZSIGCOR)
  \item Subtract bias level from reference pixels (BLEVCORR)
  \item Subtract zeroth read image (ZOFFCORR)
  \item Initialize error array, ERR (NOISCORR)
  \item Correct for detector non-linear response (NLINCORR)
  \item Subtract dark current image (DARKCORR)
  \item Compute photometric keyword values for header (PHOTCORR) 
  \item Convert to units of count rate (UNITCORR)
  \item Fit ``up the ramp'' and identify CR hits (CRCORR)
  \item Divide by flat-field image(s) and apply gain conversion (FLATCORR)
  \item Compute image statistics (No switch)
\end{enumerate}
}
Considerable additional details are in the Data Handbook, including descriptions of the calibration reference files and where to download them, and also additional steps such as conversion from 2D slitless spectra to wavelength-calibrated 1D spectra of specific targets (cf. HSTaXe in \S \ref{sec:software}).

\section{Software}
\label{sec:software}

Python packages and Jupyter notebooks are available through public repositories.\footnote{\url{https://www.stsci.edu/hst/instrumentation/wfc3/software-tools} is WFC3 specific; see also the HST repository \url{https://spacetelescope.github.io/hst_notebooks/}.}

{\bf Time-Dependent Photometry} - 
Due to the declining instrument sensitivity ($\sim0.1\%$ per year), FLT images from significantly different epochs must be converted to absolute units using the PHOTFLAM keyword (cf. IHB) prior to performing comparisons of results. Alternatively, the FLT science arrays in units of count rate (electrons/s) must be corrected for sensitivity losses before drizzling frames together. In this case, a single PHOTFLAM value may be used for the combined image. 

{\bf WFC3 Image Displayer \& Analyzer} - \textsf{This notebook provides a method to quickly display images from WFC3. This tool also allows the user to derive statistics by row or column in the image.  }

{\bf Flux Conversion Tool} - \textsf{Perform conversions between various systems of flux and magnitude using the synphot and stsynphot packages. Extrapolate an output flux at a different wavelength than the input flux, by using a spectrum defined using the same packages.}

{\bf Point Spread Function Modeling} - \textsf{Retrieve or create a PSF model for single exposures (FLTs/FLCs) and drizzled (DRZs) images. Complete basic science workflows including PSF photometry, subtraction, and decomposition. Also perform aperture photometry for comparison with PSF photometry.}

{\bf Methods for Correcting for Time-Variable Background:}

{\begin{itemize}
    \item {\bf WFC3/IR IMA Visualization Tools with an example of Time Variable Background} - \textsf{familiarizes users with the IMA file, displays individual reads and the difference between reads, and plots the cumulative signal and count rate throughout a MULTIACCUM exposure.}

    \item {\bf Correcting for Helium Line Emission Background} - \textsf{This notebook shows how to identify IR exposures with time-variable Helium (1.083 micron) line emission background, and how to correct for it using the ``flatten-ramp'' technique. This method can be used to correct images affected by a sky background that does not vary across the field of view (n.b. \underbar{not} scattered-light from the Earth's limb).}

    \item {\bf Manual Recalibration with calwf3: Turning off the IR Linear Ramp Fit} - \textsf{This notebook shows two reprocessing examples for WFC3/IR observations impacted by time-variable background (TVB).}
    
    \item {\bf Using CALWF3 to Mask Bad Reads} - \textsf{ 
    This method illustrates how to mask bad reads in the RAW image and then reprocess with CALWF3. It may also be used for rejecting anomalous reads occurring either at the beginning or at the end of an exposure, e.g. due to scattered light from the Earth's limb.}
    
    \item {\bf Manually Subtracting Bad Reads} - \textsf{
    This method illustrates how to manually subtract contaminated reads at the end of an IMA, again due to scattered light from the Earth's limb.}

\end{itemize}}

{\bf Masking Persistence in WFC3/IR Images } - \textsf{This notebook uses the persistence data products, e.g. \texttt{\_persist.fits}, to flag pixels in the FLT images. When the images are sufficiently dithered to step over the observed persistence artifacts, AstroDrizzle can exclude those flagged pixels when combining the FLT frames.}

=
{\bf HSTaXe} is a Python package for analysis of slitless spectroscopy such as from WFC3/IR.\footnote{\url{https://hstaxe.readthedocs.io/en/latest/hstaxe/examples.html} and \url{https://github.com/spacetelescope/hstaxe}} 
In 2026, HSTAxe will be superceded by {\bf Slitlessutils}, a package to analyze wide field slitless spectroscopy for WFC3 and ACS.\footnote{\url{https://slitlessutils.readthedocs.io/en/latest/}} 

\textsf{{\bf DrizzlePac} is a package to align and combine HST images.\footnote{DrizzlePac's Overview: \url{https://www.stsci.edu/scientific-community/software/drizzlepac}} To produce its drizzled (DRZ) data products, MAST uses DrizzlePac's {\bf AstroDrizzle}, which removes geometric distortion, corrects for sky background variations, flags cosmic-rays, and combines images with optional subsampling. MAST's drizzling works only for single-visit associations; for multiple visits, use {\bf TweakReg}. For manual drizzling, we suggest starting with Drizzlepac notebooks\footnote{\url{https://spacetelescope.github.io/hst_notebooks/notebooks/DrizzlePac/README.html}}.
}

\section{Acknowledgments}

The authors thank the WFC3 team throughout the years, whose dedication to this instrument has enabled the scientific returns we rely upon today. We thank Sylvia Baggett and Jennifer Mack for sharing their thorough understanding of WFC3/IR, and their historical perspectives. We thank Amanda Pagul, Mitchell Revalski, Frederick Dauphin, Mariarosa Marinelli, and Benjamin Kuhn for helpful and clarifying guidance and suggestions during their review of this report. 
The authors thank the four non-STScI reviewers of an early draft of the manuscript.

\printbibliography

\section{Appendix A: Instrument Design} \label{sec:instrument}

\begin{figure}[h]
\centering
\includegraphics[width=18cm]{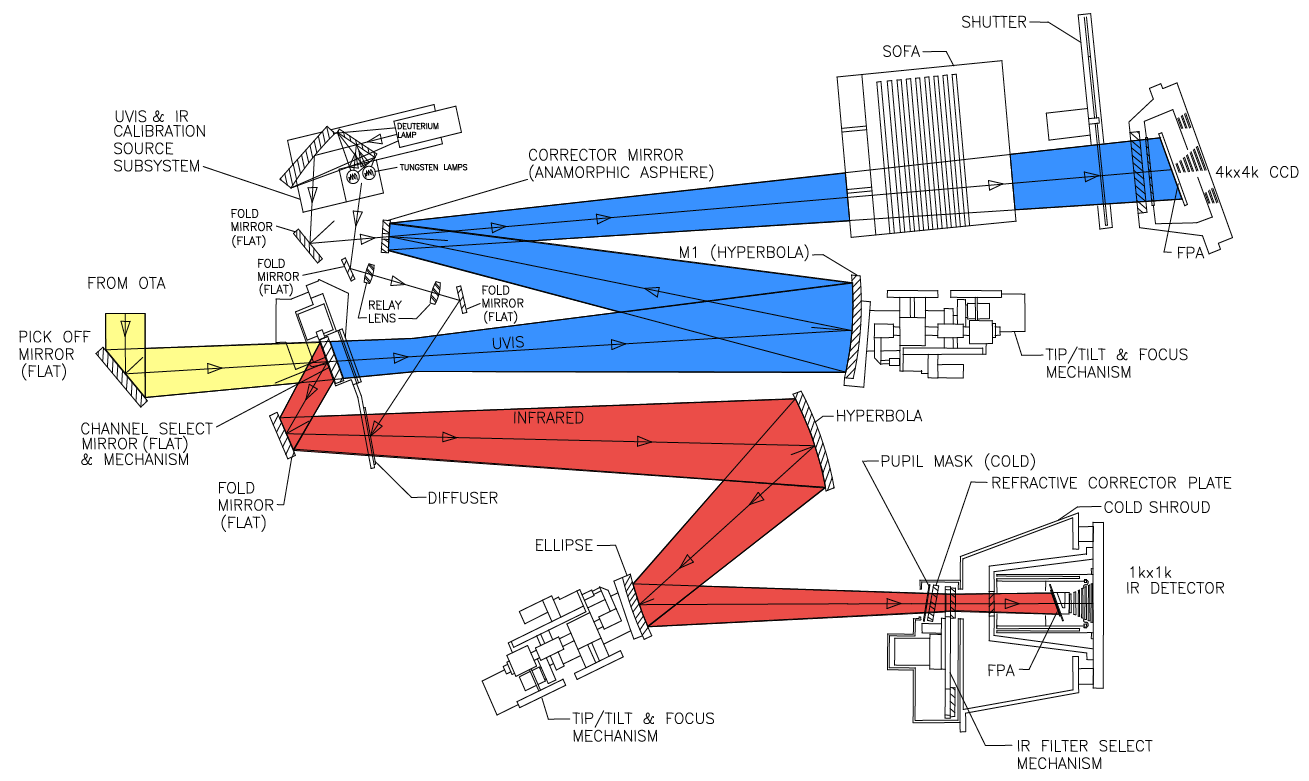}
\caption{Schematic optical layout of the WFC3 instrument. The IR channel is in red. The UVIS channel is in blue. The on-axis light from HST's secondary mirror is directed by a third pick-off mirror to the instrument (yellow). \fnihb}
\label{fig:optical-layout}
\end{figure}

 \textsf{WFC3 features two independent imaging cameras: the near-infrared channel is the topic of this document (red in Figure \ref{fig:optical-layout}); the ultraviolet / visible (UVIS) channel (blue) is not covered by this document. A mirror on the Channel Select Mechanism can then divert the light into the IR channel if desired. As a result of this design, only one channel can be used at a time, IR or UVIS. The instrument corrects HST's spherical aberration to produce a diffraction-limited image or slitless spectrum upon the IR detector, a HgCdTe array with 1024 x 1024 square pixels, each 18 microns on a side and sensitive from 800 nm to 1700 nm. Using internal lamps to illuminate the detector uniformly combined with observations of star fields, ``flat fields'' were created and are used by the CALWF3 calibration pipeline. Dark frames, i.e., those without any illumination, are observed by selecting the slot in the IR filter wheel containing an opaque disk. \fnihb}. 

Detector:

\begin{itemize}
    \item HgCdTe (Mercury Cadmium Telluride)
    \item Sensitive from 0.8 to 1.7 microns
    \item 1024 × 1024 pixels (1014 × 1014 usable)
    \item Read Noise $<$ 20 electrons r.m.s.
    \item Dark Current (includes thermal) = 0.05 electrons/s/pixel 
    \item 0.121 x 0.135 arcseconds/pixel
    \item Field of View: 123 × 136 arcseconds
    \item PSF: $\sim$2 pixels FWHM
\end{itemize}

Spectral elements:

\begin{itemize}
    \item 15 filters (5 wide, 4 medium, and 6 narrow)
    \item 2 grisms for slitless spectroscopy at low resolving power (R $\sim$150)
\end{itemize}

\clearpage
\section{Appendix B: Exposure Time Calculator Examples} \label{sec:etc}

\begin{table}[h]
  \centering
  \fontsize{11pt}{11pt}\selectfont
  \begin{tabular}{c c c c c c c c c c c}
     Time & Filter $\rightarrow$ & 105W & 110W & 125W & 140W & 160W & 098M & 127M & 139M & 153M \\
    \hline\hline
    44 s & Star, J = 22 & 718  & 1146  & 641  & 730 & 420 & 426 & 144 & 118 & 107 \\
    & Background & 19.4  & 24.6  & 19.3  & 21.3 & 17.8 & 15.2 & 10.6 & 10.1 & 10.2 \\
    & SNR & 13.57 & 19.49 & 12.29 & 13.58 & 8.50 & 8.76 & 3.25 & 2.69 & 2.44 \\
    \hline
    353 s & Star, J = 24 & 913  & 1457*  & 815  &  928 & 534 & 542 & 184 & 150 & 136 \\
                   & Background & 54.8  & 69.7*  & 54.7  & 60.3 & 50.5 & 42.9 & 30.1 & 28.7 & 28.8  \\
                   & SNR & 12.18  & 16.27*  & 10.98  & 11.72 & 7.72 & 8.48 & 3.48 & 2.90 & 2.63  \\
    \hline
    1403 s & Star, J = 25 & 1444 & 2305 & 1289 & 1468 & 846 & 858 & 290 & 237 & 215 \\
           & Background & 109.3 & 138.9 & 109.0 & 120.2 & 100.7 & 85.6 & 60.0 & 57.2 & 57.4 \\
           & SNR & 11.75 & 15.10 & 10.57 & 11.06 & 7.51 & 8.63 & 3.88 & 3.28 & 2.98 \\
    \hline
    \\
    Time & Filter $\rightarrow$ & 126N & 128N & 130N & 132N & 164N & 167N \\
    \hline\hline
    44 s & Star, J = 19 & 451 & 482 & 497 & 469 & 370 & 382 \\
        & Background           &  7.1 & 7.2 & 7.3 & 7.2 & 7.4 & 7.7 \\
        & SNR           &  9.60 & 10.19 & 10.46 & 9.94 & 8.02 & 8.24 \\
    \hline
    353 s & Star, J = 21 & 574  & 612  & 632  &  597 & 470 & 485 \\
          & Background          & 20.2  & 20.4  & 20.6  & 20.4 & 21.1 & 21.7   \\
          & SNR          & 11.07  & 11.71  & 12.03  & 11.45 & 9.18 & 9.40   \\
    \hline
    1403 s & Star, J = 22 & 908 & 969 & 1000 & 944 & 744 & 768 \\
           & Background          & 40.2 & 40.7 & 41.0 & 40.6 & 42.0 & 43.2 \\
           & SNR          & 13.96 & 14.72 & 15.10 & 14.40 & 11.46 & 11.66 \\
    \hline
  \end{tabular}
  \caption{Results from the ETC for a G2 V star for three exposure times corresponding to three sample sequences (RAPID, SPARS25, and SPARS100) each with the maximum number of samples, NSAMP = 15.
  The star's apparent brightness, specifically its J band magnitude in the Vega system, was selected for each subset in order to result in a moderate SNR. The values in the {\bf Star} row are the number of photoelectrons (\electron) attributed to the star within the r=0.2 arcsec aperture. The {\bf Background} row tabulates the corresponding background counts in the aperture in the same units (\electron). The effective read noise for 15 samples and an area of 7.66 pixels equals 41.3 \electron\ in all cases, so it is not shown.
  From these data, graphs like Figure \ref{fig:error_vs_mag} can be constructed for specific filters and exposure times by applying the powerlaw indices of each component, picking an appropriate value for the systematic noise floor, and combining values in quadrature to calculate the SNR, which should match the value tabulated at the indicated stellar J magnitude (\S \ref{sec:error}). Values specific to Figure \ref{fig:error_vs_mag} are indicated with asterisks; for those, the SNR value of 16.27 equals the reciprocal of the square root of the sum of the squares of three values: 1/$\sqrt{1457}$, 69.7/1457, and 41.3/1457. Note that the latter calculation neglects the noise floor; to include it, sum the squares of {\it four} values, with the fourth value equaling the noise floor, typically 0.008 (cf. \S \ref{sec:error}) which results in a SNR = 16.13.}
  \label{tab:etc}
\end{table}
\newpage
\section{Appendix C: Glossary of Uncommon Acronyms} \label{sec:glossary}

\begin{multicols}{2}
{\bf ABMAG} ``ABsolute'' MAGnitude
\\{\bf ACS} Advanced Camera for Surveys
\\{\bf ALMA} Atacama Large Millimeter Array
\\{\bf CR} Cosmic Ray
\\{\bf DASH} Drift And Shift
\\{\bf DHB} Data Handbook
\\{\bf DQ} Data Quality
\\{\bf DRZ} drizzled data file
\\{\bf ERR} error array (FITS)
\\{\bf ETC} Exposure Time Calculator
\\{\bf FITS} Flexible Image Transport System
\\{\bf FLT} flattened data file
\\{\bf GUI} Graphical User Interface
\\{\bf HAP} Hubble Advanced Products
\\{\bf IHB} Instrument Handbook
\\{\bf IMA} Intermediate MultiAccum data file
\\{\bf ISR} Instrument Science Report
\\{\bf MAST} Mikulski Archive for Space Telescopes
\\{\bf NICMOS} Near Infrared Camera MultiObject Spectrograph
\\{\bf NSAMP} Number of SAMPles (FITS keyword)
\\{\bf PAM} Pixel Area Map
\\{\bf PCS} Pointing Control System
\\{\bf PHOTFLAM} Synphot-generated keyword
\\{\bf PSF} Point Spread Function
\\{\bf RA} Right Ascension
\\{\bf RAW} raw data file
\\{\bf SAMP} sample (FITS keyword)
\\{\bf SCI} science data file
\\{\bf SDSS} Sloan Digital Sky Survey
\\{\bf SNR} Signal to Noise Ratio
\\{\bf SPARS} Sparse sampling sequence
\\{\bf TVB} Time Variable Background
\\{\bf UV} UltraViolet
\\{\bf UVIS} Ultraviolet-VISible
\\{\bf VLA} Very Large Array
\\{\bf WCS} World Coordinate System
\end{multicols}

\end{document}

%% file: abbreviations.tex

\newcommand{\fnihb}{[IHB]}
\newcommand{\fndhb}{[DHB]}

\newcommand{\tess}{{\it TESS}}
\newcommand{\jwst}{{\it JWST}}
\newcommand{\kepler}{{\it Kepler}}
\newcommand{\ktwo}{{K2}}
\newcommand{\hst}{{\it HST}}
\newcommand{\swift}{{\it Swift}}
\newcommand{\integral}{{\it INTEGRAL}}
\newcommand{\nustar}{{\it NuSTAR}}
\newcommand{\fermi}{{\it Fermi}}
\newcommand{\msun}{$M_{\odot}$}
\newcommand{\rsun}{$R_{\odot}$}
\newcommand{\lsun}{$L_{\odot}$}
\newcommand{\re}{$R_{\oplus}$}
\newcommand{\me}{$M_{\oplus}$}
\newcommand{\kms}{km~s$^{-1}$}
\newcommand{\fluxcgs}{ergs~s$^{-1}$~cm$^{-2}$}
\newcommand{\lumcgs}{ergs~s$^{-1}$}
\newcommand{\rj}{$R_{\textrm{\scriptsize Jup}}$}
\newcommand{\mj}{$M_{\textrm{\scriptsize Jup}}$}
\newcommand{\electron}{e$^{-}$}

\newcommand\aj{{AJ}}
\newcommand\araa{{ARA\&A}}
\newcommand\apj{{ApJ}}
\newcommand\apjl{{ApJL}}     
\newcommand\apjs{{ApJS}}
\newcommand\ao{{ApOpt}}
\newcommand\apss{{Ap\&SS}}
\newcommand\aap{{A\&A}}
\newcommand\aapr{{A\&A~Rv}}
\newcommand\aaps{{A\&AS}}
\newcommand\azh{{AZh}}
\newcommand\baas{{BAAS}}
\newcommand\icarus{{Icarus}}
\newcommand\jaavso{{JAAVSO}}  
\newcommand\jrasc{{JRASC}}
\newcommand\memras{{MmRAS}}
\newcommand\mnras{{MNRAS}}
\newcommand\pra{{PhRvA}}
\newcommand\prb{{PhRvB}}
\newcommand\prc{{PhRvC}}
\newcommand\prd{{PhRvD}}
\newcommand\pre{{PhRvE}}
\newcommand\prl{{PhRvL}}
\newcommand\pasp{{PASP}}
\newcommand\pasj{{PASJ}}
\newcommand\qjras{{QJRAS}}
\newcommand\skytel{{S\&T}}
\newcommand\solphys{{SoPh}}
\newcommand\sovast{{Soviet~Ast.}}
\newcommand\ssr{{SSRv}}
\newcommand\zap{{ZA}}
\newcommand\nat{{Nature}}
\newcommand\iaucirc{{IAUC}}
\newcommand\aplett{{Astrophys.~Lett.}}
\newcommand\apspr{{Astrophys.~Space~Phys.~Res.}}
\newcommand\bain{{BAN}}
\newcommand\fcp{{FCPh}}
\newcommand\gca{{GeoCoA}}
\newcommand\grl{{Geophys.~Res.~Lett.}}
\newcommand\jcp{{JChPh}}
\newcommand\jgr{{J.~Geophys.~Res.}}
\newcommand\jqsrt{{JQSRT}}
\newcommand\memsai{{MmSAI}}
\newcommand\nphysa{{NuPhA}}
\newcommand\physrep{{PhR}}
\newcommand\physscr{{PhyS}}
\newcommand\planss{{Planet.~Space~Sci.}}
\newcommand\procspie{{Proc.~SPIE}}

\newcommand\actaa{{AcA}}
\newcommand\caa{{ChA\&A}}
\newcommand\cjaa{{ChJA\&A}}
\newcommand\jcap{{JCAP}}
\newcommand\na{{NewA}}
\newcommand\nar{{NewAR}}
\newcommand\pasa{{PASA}}
\newcommand\rmxaa{{RMxAA}}

\newcommand\maps{{M\&PS}}
\newcommand\aas{{AAS Meeting Abstracts}}
\newcommand\dps{{AAS/DPS Meeting Abstracts}}

\let\astap=\aap 
\let\apjlett=\apjl 
\let\apjsupp=\apjs 
\let\applopt=\ao